\documentclass[12pt]{article}
\usepackage{graphicx}
\usepackage{cite}
\newcommand{\be}{\begin{equation}}
\newcommand{\ee}{\end{equation}}
\newcommand{\beq}{\begin{eqnarray}}
\newcommand{\eeq}{\end{eqnarray}}
\newcommand{\ba}{\begin{array}}
\newcommand{\ea}{\end{array}}

\date{}

\begin{document}
\title{
\large \bf Legendre Analysis of Hadronic Reactions}
\author{ Ya.I.~Azimov$\,^1$ }

\maketitle
\vspace{-7mm}
\centerline{\small $~~~~~~~~~^1$\it {Petersburg Nuclear Physics Institute,
NRC Kurchatov Institute, }}
\centerline{\small \it {    Gatchina, 188300, Russia}}

\vspace{-4mm}
\begin{abstract}
Expansions over Legendre functions are suggested as a model-independent
way of compact presentation of modern precise and high-statistics data for
two-hadron reactions. Some properties of the expansions are described.
\end{abstract}

\vspace{5mm}

Modern detectors combined with modern accelerator facilities are capable
to provide tremendous sets of experimental data. Here the problem arises,
how to present those numerous detailed data. For relatively simple cases
of $2\to2$ reactions, there are two popular methods. The data are usually
presented either as pictures with tens of small panels showing angular
distributions for different fixed energies or as similar multi-panel pictures
showing excitation functions, \textit{i.e.}, energy distributions at
different fixed angles (see \textit{e.g.}, Refs.\cite{CBC,CLAS,A2}). Both
approaches may be used, of course, to check various models, but are not
practical for any direct extraction of information. They both are very bulky
and non-visual, especially if one would try to present the whole data-set.

As an alternative, we have suggested to expand experimental data into series
of the Legendre functions. For unpolarized differential cross sections, it
means decomposition
\be
\frac{d\sigma}{d z}(E,z)=\sum_{J=0}^{\infty}\,A_J(E)\,P_J(z)\,,
\label{decomp}
\ee
where  $E$ is the c.m. energy, $z=\cos\theta$, and $\theta$ is the polar c.m.
angle. Formally, this series is infinite. However, in real situations only
some finite number of the Legendre coefficients $A_J(E)$ are efficient, since
higher coefficients have errors that exceed fitted values of those coefficients.
This was illustrated in Refs.\cite{CBC,A2} for photoproduction reactions
$\gamma p\to\eta p$ and $\gamma p\to\pi^0p$ respectively. As a result, the whole
set of data appear to be presented in a compact form of energy dependencies for
several Legendre coefficients.

Let us briefly discuss properties of the coefficients $A_J\,$. Each of them is
a sum, formally infinite, of terms bilinear in respect to partial-wave amplitudes.
Their angular momenta, $j_1$ and $j_2\,$, satisfy the familiar conditions
\be
|j_1-j_2|\leq J \leq j_1+j_2\,.
\label{3j}
\ee
The summations over $j_1$ and $j_2$ should go to infinity. However the partial-wave
amplitudes decrease at high $j$ (exponentially in asymptotics), so, when accounting
for experimental uncertainties, one needs only a finite number of summands.

The Legendre coefficients have also other, less evident properties. For parity conserving
reactions, some of those properties are related to partial-wave amplitudes for states
of definite parity. In the unpolarized cross section the positive- and negative-parity
amplitudes always appear symmetrically (that is why the unpolarized cross section by
itself does not allow to determine the parity of a particular state). It is not quite so
for $A_J\,$. One can show that the Legendre coefficients provide specific discrimination
of parities: $A_J$ with odd $J$ contain only interferences of states with opposite parities,
while $A_J$ with even $J$ contain only interferences of states with the same parities,
positive or negative. And, of course, only the even-$J$ coefficients may contain squares
of absolute values of various partial-wave amplitudes.

It is interesting to discuss how resonances reveal themselves in the Legendre coefficients.
Let us consider the simplest idealized case of a pronounced resonance with spin $S$ over
very small (negligible) background. Then we practically have only one partial-wave amplitude
with $j=S\,$. According to frequent belief, cross section in such a case contains the resonance
contributions into the Legendre harmonics with $J$ up to $2S\,$. But this is not quite true. As
explained above, the resonance contributions (with the same parity and without background!)
may appear not in every $A_J\,$, only in coefficients with \textit{even} $J$-values. Moreover,
for fermionic resonances having half-integer $S\,$, the value of $2S$ is odd, hence $A_J$
with $J=2S$ does \textit{not} contain the resonance contribution! Of course, a boson resonance
having integer $S$ does contribute to $A_J$ with $J=2S\,$.

The situation changes if there are background contributions, especially if they may have both
positive and negative parities. Due to interferences resonance-background, the resonance signals
become present in every $A_J\,$, with either even or odd value of $J\,$. Moreover, if the resonance
is sufficiently intensive, as \textit{e.g.}, $\Delta(1232)$, the resonance interference can work
as an amplifier (see examples in Ref.\cite{YaA}). It can enhance contributions of very small
partial-wave amplitudes, which, by themselves, are not seen \textit{vs.} experimental errors.
Here we encounter an interesting point. If parity of the resonance is known, parity of interfering
states becomes also known, due to the discrimination of parities between even and odd values 
of $J\,$, as explained above.

The Legendre decomposition may be applied also to polarization variables, more exactly, to polarization
variables multiplied by the unpolarized cross section. However, instead of Legendre polynomials, one
should generally use other functions, associated Legendre functions or even Wigner $d$-harmonics. Such
approach was applied to experimental data on the beam asymmetry~\cite{CLAS}, which was  expanded into
a series over the associated functions $P_J^2(z)\,$.

We can summarize the above statements in the following way:
\begin{itemize}

\item{Legendre expansions provide a model-independent approach suitable for presentation of modern
detailed (high-precision and high-statistics) data for two-hadron reactions;}
\item{This approach is applicable both to cross sections and to polarization variables; it is much
more compact than traditional methods, at least, at not very high energies;}
\item{The Legendre coefficients reveal specific correlations and interferences between states of
definite parities;}
\item{Due to interference with resonances, high-momentum Legendre coefficients open a window to study
higher partial-wave amplitudes, which are out-of-reach in any other ways.}

\end{itemize}

More detailed discussion of the Legendte expansions, including proofs of various properties stated above as
well as examples of physical information which may be obtained by their applications, will be given in a
separate paper~\cite{abs}.

This work was supported by the Russian Science Foundation (Grant No.14-22-00281).

\end{document}